\documentstyle[12pt,a4,epsf]{article}
\def\be{\begin{equation}}
\def\ee{\end{equation}}
\def\beq{\begin{eqnarray}}
\def\eeq{\end{eqnarray}}
\def\lsim{\:\raisebox{-0.9ex}{$\stackrel{<}{\sim}$}\:}
\begin{document}
\begin{flushright}
TIFR/TH/96-54
\end{flushright}
\medskip
\begin{center}
{\large\underbar{\bf Zero-temperature Hysteresis in Random-field
Ising}} \\[.5cm]
{\large\underbar{\bf Model on a Bethe Lattice}} \\[1cm]
Deepak Dhar$^{1,3}$, Prabodh Shukla$^{2,3}$, and James P. Sethna$^4$ 
\end{center}
%\vspace{1cm}
\begin{enumerate}
\item[{1}] Tata Institute of Fundamental Research, Homi Bhabha Road,
\\ Mumbai 400 005, INDIA \\ e-mail: ddhar@theory.tifr.res.in
\item[{2}] Physics Department, North Eastern Hill University, \\
Shillong 793 022, INDIA \\e-mail:shukla@nehus.ren.nic.in
\item[{3}] International Centre for Theoretical Physics, \\
P.O. Box 586, Trieste 34100, ITALY
\item[{4}] Laboratory of Atomic and Solid State Physics, \\
Cornell University, Ithaca, NY 14853-2501, USA 
\\ e-mail:sethna@lassp.cornell.edu
\end{enumerate}
%\vspace{1cm}
\begin{center}
\underbar{\bf ABSTRACT}
\end{center}
%\medskip

We consider the single-spin-flip dynamics of the random-field Ising model
on a Bethe lattice at zero temperature in the presence of a uniform
external field.  We determine the average magnetization as the external
field is varied from $-\infty$ to $+\infty$ by setting up the
self-consistent field equations, which we show are exact in this case.  
The qualitative behavior of magnetization as a function of
the external field unexpectedly depends on the coordination number $z$ of
the Bethe lattice. For $z = 3$, with a gaussian distribution of the
quenched random fields, we find no jump in magnetization for any non-zero
strength of disorder. For $z\ge4$, for weak disorder the magnetization
shows a jump discontinuity as a function of the external uniform field,
which disappears for a larger variance of the quenched field. 
We determine
exactly the critical point separating smooth hysteresis curves from those
with a jump. We have checked our results by Monte Carlo simulations of the
model on 3- and 4- coordinated random graphs, which for large system sizes
give the same results as on the Bethe lattice, but avoid surface effects
altogether.  \medskip

 \newpage

\begin{center} {\large \underbar{\bf I. Introduction}} \end{center}
\medskip

Recently, a simple model has been introduced [1] for hysteresis in
magnets, which incorporates interesting effects like the return-point
memory and Barkhausen noise [2].  In this model, Ising spins with a
quenched random field at each site evolve by a zero-temperature
single-spin-flip dynamics.  The authors argued that in this model, if the
external field is increased slowly, the steady-state magnetization as a
function of the field has a jump discontinuity at some critical value of
the field for a small disorder, but is a continuous function with no jump
discontinuity for large disorder. This picture was supported by numerical
simulations of the model on hypercubic lattices in two and higher
dimensions.  Subsequent work [3] studied in detail the transition from
jump to no-jump in magnetization at a critical value of the gaussian
disorder, and observed scaling behavior in the neighbourhood of this
critical disorder.  However, the exact solution of this model in one 
dimension does not show a jump discontinuity for gaussian disorder
neither  for the ferromagnetic nor for antiferromagnetic exchange
couplings[4]. 

In this paper, we extend the treatment of [4] to study this hysteresis
model on a Bethe lattice.  The Bethe lattice of coordination $z$ is the
formal infinite-size limit of a branching tree (the Cayley tree) where
each spin has $z$ nearest neighbors, and the statistical averages are
calculated away from the `surface'of the lattice.  For $z=3$, the lowest
non-trivial coordination, we find no jump in magnetization for any nonzero
disorder, if the quenched random fields have a gaussian distribution ---
just as in the one-dimensional case. For coordination $z>4$, we find there
is a non-zero critical disorder where the macroscopic jump in
magnetization in the hysteresis loop first disappears. 

This is very
surprising, as in all the models studied on the Bethe lattice, the
qualitative behavior of the solution has been found to be independent of
the coordination number ( so long as it is greater than 2). In particular,
a Bethe lattice with finite coordination number $z$ has the same critical
behavior as the mean-field theory, which corresponds to the limit of large
coordination number $z$ and coupling constant scaling as $1/z$.  The
reason why this unusual dependence on $z$ shows up in this problem is not
yet understood. 

Our treatment is based on setting up self-consistent equations for some
nearest neighbour correlation function in the problem [5].  We can show
that these self-consistent equations are exact in this case, though we are
not aware of a rigorous proof that this happens in general for a Bethe
lattice in the presence of quenched disorder.  In fact, the presence of
the disorder usually renders the problem analytically intractable.  For
example, for the Ising spin-glass problem on a Bethe lattice with random
$\pm J$ bonds, it has not been possible to determine exactly even the zero
temperature quantities like the ground state energy or the ground state
entropy [6,7]. However, the qualitative behavior of the system near the
thermal critical point seems fairly well understood [8]. 

The plan of this paper is as follows:  In Section II, we define the model
precisely.  In Section III, we set up recursion relations on a Cayley tree
for conditional probabilities that the spin at a given site at height $r$
from the boundary is down, given that the spin on its parent, ``upward''
neighbor on the tree is down. We are interested in the intensive
quantities, such as magnetization or energy density on the tree far away
from the boundary.  These turn out to be independent of details of the
boundary conditions, and we take this as the definition of Bethe
approximation in our case.  We obtain an explicit expression for
magnetization as a function of external field for arbitrary distribution
of the quenched random fields. In Section IV, we describe a method to
simulate spin systems on the Bethe lattice that is computationally
efficient, and does not suffer from surface effects.  We use this method
to check the validity of our self-consistent equations for the case of
gaussian and rectangular distributions of quenched random fields.  The
agreement is found to be excellent. Section V contains some concluding
remarks.  \bigskip

\begin{center} {\large\underbar{\bf II. The Model}} \end{center} \medskip

\nobreak We consider a lattice of $N$ sites.  Each site is labeled by an
integer $i = 1$ to $N$, and carries an Ising spin $S_i$ $(S_i = \pm 1)$
which interacts with a finite number $z$ of neighbouring spins with a
ferromagnetic interaction $J$.  There is a uniform magnetic field $h$
which is applied externally.  In addition, at each site $i$, there is a
local quenched random field $h_i$.  The fields $\{h_i\}$ are assumed to be
independent identically distributed random variables with a continuous
probability distribution $p(h)$.  The system is described by the
Hamiltonian \be H = -J \sum_{\langle ij\rangle} S_i S_j - \sum_i h_i S_i -
h \sum_i S_i. \ee The zero-temperature single-spin-flip Metropolis-Glauber
dynamics [9] is specified by the transition rates \beq {\rm Rate} \ [S_i
\rightarrow -S_i] &=& \Gamma, \ {\rm if} \ \Delta E \leq 0; \nonumber
\\[2mm] &=& 0, \ {\rm otherwise}; \eeq where $\Delta E$ is the change of
energy of the system as a result of the spin-flip.  We shall be interested
in long time scales $\gg \Gamma^{-1}$.  In this limit, the dynamical rule
simplifies to the following: Choose a spin at random, and flip it only if
this process would lower the energy. Repeat till a stable configuration is
obtained. 

The problem of hysteresis which we address here is as follows: Start with
a sufficiently large negative applied field $h$, so that in the stable
configuration all spins are down ($S_i = -1$, for all $i$) and increase
the field slowly.  At some value of $h$, the local field $\ell_i$ at some
site $i$, defined by \be \ell_i = J \sum_j S_j + h_i + h \ee will become
positive, and this spin would flip up.  [The summation in (3) is over all
the neighbors $j$ of $i$].  This changes the effective field at the
neighbors, and some of them may flip up, and so on, causing an avalanche
of flipped spins.  We determine the total magnetization when the avalanche
has stopped.  Then we raise the applied field a bit more, and determine
the magnetization in the stable state again.  The process is continued
until all the spins flip up.
This generates the lower half of the
hysteresis loop (plot of magnetization $m(h)$ versus $h$) in the situation
where the applied field is varied very slowly, or equivalently, when the
spins relax infinitely fast.  The upper half of the hysteresis loop
$m_u(h)$ is obtained when the field $h$ is decreased from $+\infty$ to
$-\infty$.  This is related to the lower half of the loop $m_{\ell}(h)$ by
symmetry 
\be 
m_u(h) = -m_{\ell}(-h). 
\ee

This corresponds to the zero frequency limit
of a driving field oscillating sinusoidally in time with frequency
$\omega$.  Note that the limit of $\omega \rightarrow 0$ is taken after
the limit temperature T $\rightarrow 0$. If the limits are taken in the
reverse order, the area of the hysteresis loop goes to zero as $\omega $
goes to zero for all non-zero T.

An important feature of the above dynamics for ferromagnetic couplings is
that if we start with any stable configuration, and then increase the
external field and allow the system to relax, then in the relaxation
process no spin flips more than once.  Furthermore, the final stable
configuration is the same whatever the order in which unstable spins are
flipped.  This property is called the `no passing property' [10], and
greatly simplifies the analysis.  \bigskip

\begin{center} {\large\underbar{\bf III. Recursion Relations on the Cayley
Tree}} \end{center} \medskip

\nobreak The standard approach for solving statistical mechanics problems
on the Bethe lattice is to consider the problem on a Cayley tree, and
consider behavior deep inside the tree, i.e. far from the boundaries of
the tree [11,12].  If suitable care is taken to remove the effects of the
boundary, all correlation functions deep inside the Cayley tree (say for
the Ising model with an external field) are found to be the same as in the
Bethe approximation. Thus, we may say that the Bethe lattice is the {\it
deep interior} part of the Cayley tree.  Here we shall use this approach.
In Section IV, a different approach is presented. 

Consider a Cayley tree of height $n$.  Each site of the tree has
coordination number $z$, except the boundary sites which have coordination
number $1$.  The level $n$ consists of only one site $O$, called the
central site. For $r \geq 1$ the level $(n-r)$ has exactly
$z\cdot(z-1)^{r-1}$ sites [Fig. 1]. 

We start with the external field $h$ large and negative, so that ground
state of the system is with all spins down.  Now, increase the external
field to a finite value $h$, and flip up any spin for which the net local
field is positive.  As the same final stable configuration is attained,
whatever the order in which spins are relaxed, we may start by first
relaxing spins of level 1.  Then we relax spins of level 2, then of level
3, and so on.  If a spin at level $r$ is flipped up, we check all its
descendents again for possible upward flips. 

Let $P_r$ be the conditional probability that a randomly chosen spin at
level $r$ is upturned in this scheme, given that its parent spin at level
$(r+1)$ is kept down, and the spin and all its descendent spins are
relaxed as far as possible.  Let $S_r$ be the spin at level $r$. We relax
all the descendent spins of $S_r$ first, keeping $S_r$ down.  In this
process, each of the $z-1$ direct descendents of $S_r$ at level $(r-1)$ is
independently flipped up with probability $P_{r-1}$.  Hence the
probabilities that $z-1,z-2,\dots,0$ of the children of $S_r$ are flipped
up in this relaxation process are $P^{z-1}_{r-1}$, $(z-1)
P^{z-2}_{r-1}(1-P_{r-1})$, \dots, $(1-P_{r-1})^{z-1}$ respectively. 
Consider the case where $s$ of the children are up:  since the parent
neighbor remains down for this part of the calculation, the net number of
down neighbors is $z - 2s$, and hence, the spin $S_r$ will flip up if the
local field at this site exceeds $(z-2s)J-h$.  Let $p_s(h)$ denote the
probability that the local field at a randomly chosen site is large enough
so that the spin will flip up if $s$ of its children are up, and the
uniform field is $h$.  Clearly \beq p_s(h) &=& {\rm Prob \ that \ local \
field} \geq -h + zJ - 2sJ \nonumber \\[2mm] &=& \int^\infty_{-h+(z-2s)J}
p(h_i) dh_i. \eeq Then it is easily seen, e.g. for $z=3$ that \be
(z=3):~~P_r = P^2_{r-1} p_2 (h) + 2P_{r-1} (1 - P_{r-1}) p_1(h) +
(1-P_{r-1})^2 p_0(h). \ee Given a value of $h$, we determine the
quantities $p_s(h)$. Then using Eq. (6), and the initial condition $P_1 =
p_1 (h)$, we can recursively determine $P_r$ for all $r \geq 2$.  For
large $r \ll n$, $P_r$ tend to a fixed point $P^\star$ given by the
self-consistent equation \be P^\star = \sum^{z-1}_{r=0}
\left({}^{z-1}_r\right) P^{\star r} (1-P^\star)^{z-1-r} p_r(h) \ee

This is a polynomial equation in $P^\star$, which can be solved in terms
of $\left\{p_s(h)\right\}$.  Finally, for the central site $O$ at level
$n$, there are $z$ children, and a similar argument gives \be {\rm Prob}
(S_O = +1) = \sum^z_{r=0} \left({}^z_r\right) P^{\star r}
(1-P^\star)^{z-r} p_r(h) \ee

Substituting the value of $P^\star$, from Eq. (7), we determine the
probability that this spin $S_O$ is up, and hence the average
magnetization at this site. 

The arguments above do not require that all the z descendent subtrees of
$O$ be of equal height.  So long as $O$ is sufficiently far from the
boundary, we get the same conditional probability $P^\star$, and hence the
same value of magnetization.  This proves that all sites `deep inside' the
tree have the same average magnetization. 

\bigskip

\begin{center} {\large\underbar{IV. Simulations}} \end{center} \medskip

\nobreak The derivation of our self-consistent equations assumes the
existence of a unique thermodynamic state deep within the Cayley tree
which is independent of boundary conditions. While this is quite
plausible, uniqueness of the Gibbs state has been proved so far for the
RFIM only in one dimension, and only for a bivariate distribution of the
quenched field, and nonzero temperatures [13].  It seems desirable to have
a direct check of these equations by numerical simulations which do not
involve making any assumptions about the thermodynamic state.  

While the
procedure of the previous section treating the Bethe lattice as sites deep
inside the Cayley tree is well known and conceptually simple, it is not
suited for numerical simulations.  Most of the sites of the Cayley tree
are within a short distance from the surface, and cannot be used for
averaging.  Since the `bulk' forms a negligible fraction of all possible
sites, special care has to be taken to subtract the surface contribution.
For our simulations, we used a different technique that is computationally
efficient and gets rid of surface effects altogether.  This technique has
been used earlier to study spin systems on random graphs by Monte Carlo
simulations [14]. 

Our simulation algorithm involves construction of a random graph having
$N$ sites such that each site has exactly $z$ neighbors.  The precise
algorithm we used was as follows: Label the $N$ sites by integers from 1
to $N$.  We shall assume $N$ is even in the following.  Connect site $i$
to site $(i+1)$ for all $i$.  Site $N$ is connected to site 1. This gives
us a ring of $N$ sites.  Now construct $(z-2)$ independent random pairing
of $N$ sites into $N/2$ pairs, and add a bond for each of the paired
sites.  Thus, we get a graph in which each site has coordination number
$z$ [Fig. 2]. 

In this construction, all sites are on same footing, and there is no
`surface'.  Unlike Cayley tree, this graph has loops.  However it is easy
to see that there are typically very few small loops.  For example, for
$z=3$ the probability that sites $i$, $(i+1)$ and $(i+2)$ form a loop of
length 3 is the probability that site $i$ is paired with $(i+2)$, and
equals $1/(N-1)$.  Thus the expected number of loops of size 3 in a $z=3$
graph of $N$ sites tends to 1 for large $N$.  Similarly, it can be shown
that the expected number of loops of length 4 is 2 for large $N$.  In
general, the average number of loops of length $\ell$ increases as
$\lambda^\ell$ with $\lambda=z-1$ for the random graph with coordination
number $z$, and is a negligible fraction of all sites belong to any loop
of length $\leq \ell$ for $\ell \ll \log N / \log \lambda $ [15]. 

If the smallest loop going through a given site is of length $\leq
(2d+1)$, then it follows that up to a distance $d$ from that site, the
lattice looks like a Bethe lattice.  Hence our random lattice would look
like a Bethe lattice for $z=3$ at almost all sites for a distance $\lsim
\log_2 N$.  This, in turn, can be shown to imply that in the thermodynamic
limit $N \rightarrow \infty$, the free energy per site on our lattice for
classical statistical mechanical models with short range interactions (say
nearest neighbor only) are the same as in the Bethe-Peierls approximation. 

In our simulations, we used $N = 10^6$. We used simple scanning to decide
which spins to be flipped at the next time step.  The dotted lines in
Figs. 3 and 4 show the results of a simulation for $z=3$ for quenched
gaussian random fields with mean 0 and variance $\sigma=1$ and $\sigma=3$
respectively.  The lower and upper halves of the hysteresis loop were
obtained separately in the simulation.  Also shown in the figures are the
results of solution of Eqs. (7-8).  The statistical errors of the
simulation are quite small.  Different runs, with different realizations
of quenched fields give results which are indistinguishable at the scale
of the graph.  The agreement with the theoretical calculation is
excellent. For much smaller values of disorder $\sigma \lsim .1$, the
hysteresis loops are very approximately rectangular.  In this case, the
value of coercive field is governed by the largest realized value of
quenched local field, which shows noticeable sample to sample
fluctuations. As noted above, for $z=3$ the hysteresis loop is smooth for
all values of the disorder greater than zero: the quadratic equation (6)
has only one stable solution. 

For $z=4$, we do find a transition.  At small disorder, the hysteresis
loop has a jump: one large event flips a large, finite fraction of the
spins in the thermodynamic limit.  Figure 5 shows the analytical and
simulation results for $\sigma=1.75$; there is a large jump in the
magnetization at $h=1.0037$.  The critical value of the disorder is very
slightly larger than this ($\sigma_c^{(z=4)}=1.78126$) and above the
critical disorder the hysteresis loop is smooth.  The critical field
$h_c=1$ at the critical disorder for $z=4$ and we conjecture for larger
$z$ as well. This simple result follows from the observation that at
$h=1$, $P=1/2$ is always a fixed point for Eq. (7) for all $z$:
$\sigma_c(z)$ may be determined by making this fixed point a double 
root. 
This makes the transition a traditional saddle-node transition (the lower
branch merges with an ``unstable'' branch of the self-consistent
equation). The critical exponents are thus the same as that for the
infinite-range mean-field model (which also undergoes a saddle-node
transition in its self-consistent equation). We have checked that this
same pattern also occurs for $z=5$ (where it gives $\sigma_c=2.58201$),
and conjecture that it gives the correct critical point for all 
$z>3$; in
$z=3$ the coalescence between the stable and unstable branches of the
$M(h)$ curves never occurs. 

\bigskip

\begin{center} {\large\underbar{\bf V. Discussion}} \end{center} \medskip

\nobreak It is natural to compare the zero temperature hysteresis on the
Bethe lattice with the corresponding infinite-range mean field result
obtained in the limit of large coordination number when the ferromagnetic
coupling is taken to be $J/N$, the same for all pairs of sites. In this
case the mean-field solution is given by [1] \be m = erf\left[{Jm
+ h \over \sqrt{2\sigma^2}}\right]. \ee For $\sigma < \sigma_c =
\sqrt{2/\pi}$, the above equation has two solutions $m^\star_{\ell} (h)$
and $m^\star_u (h)$ which are related to each other by the symmetry
$m^\star_{\ell} (-h) = -m^\star_u (h)$.  These correspond to the two
halves of the hysteresis loop for increasing and decreasing field
respectively.  

For $\sigma > \sigma_c$, Eq. (9) has a single valued real solution
$m^\star (h)$ which is an odd function of $h$.  Thus there is no
hysteresis for $\sigma > \sigma_c$.  The remanence goes to zero
continuously as $\sigma$ tends to $\sigma_c$ from below.  For $\sigma <
\sigma_c$, there is a discontinuity in magnetization at a critical field
$h_c$.  The value of $h_c$, and the magnitude of jump in magnetization
both tend to zero continuously as $\sigma \rightarrow \sigma_c$. This lack
of hysteresis for $\sigma>\sigma_c$ is an artifact of the hard-spin
infinite-range model: our Bethe lattice has hysteresis at all values of
$\sigma$ (as does the infinite-range model with continuous spins in a
double-well potential [1]). 

For $z>3$ the behavior as one approaches $\sigma_c$ from below is similar
to the infinite-range model, and the corresponding critical exponents
$\beta$ and $\delta$ will be the same [1].

So far, we have discussed the case when the quenched random fields have an
unbounded distribution.  For bounded distributions of the quenched random
field, one can get jumps in magnetization even for $z=3$. Consider, for
example, the case when $\{h_i\}$ have a uniform rectangular distribution
between $-h_{max}$ and $+h_{max}$[16]. If we start with all spins down,
and increase field slowly, clearly nothing happens for $h < 3J - h_{max}$. 
If $h$ exceeds this value, then the spin with the largest value of $h_i$
will flip up.  If $h_{max} < J$, this will make the net local field at the
neighbors positive.  These spins will flip up, which in turn flips their
neighbors, and so on. Thus, for $h_{max} < J$, the magnetization $m(h)$
jumps discontinuously from $-1$ to $+1$, as $h$ cross $3J - h_{max}$ [17].
If $J < h_{max} < 2J$, one can show that same thing occurs as the system
is, on the average, unstable for creating such a `nucleus' of up spins.
However, for $h_{max} > 2J$, this particular instability is absent, and
the magnetization is a continuous function of $h$. Note that the
magnetization jump goes {\it discontinuously} to zero.  If the
distribution of quenched fields $p(h_i)$ has delta functions, in addition
to a continuous part, clearly this will lead to discontinuities in the
$m(h)$ curve.  Any other singularities of $p(h_i)$, say at $h_i = \alpha$
lead to singularities in $m(h)$ for $h=\alpha \pm 3J$, $\alpha \pm J$. 

An interesting open question, which we have not been able to answer so far
is to characterize all possible `metastable' states on the Bethe lattice. 
Are all of these obtainable as solutions of self-consistent equations of
the type discussed above?  For example, can one calculate the
magnetization when the external field is first increased monotonically
from $-\infty$ to a value $H_1$, and then reduced to a value $H_2 < H_1$? 
Further study of such questions would perhaps help in our understanding of
the more general question of hysteretic dynamics of systems with many
metastable states. 

Acknowledgments: We would like to thank Drs. M. Barma and S. N. Majumdar
for critically reading the manuscript, Dr. D. A. Johnston for bringing his
earlier work on Monte Carlo simulations of spin systems on random graphs
to our notice, and A. A. Ali and Matt Kuntz for help in preparing the
figures. 

\newpage

\begin{center} {\large\underbar{\bf References}} \end{center} \bigskip

\begin{enumerate} \item[{[1]}] J. P. Sethna, K. Dahmen, S. Kartha, J. A.
Krumhansl, B. W. Roberts and J. D. Shore, Phys. Rev. Lett. 
\underbar{70},
3347 (1993); O. Perkovi\'c, K. Dahmen and J. P. Sethna, Phys. Rev. Lett.
\underbar{75}, 4528 (1995); K. Dahmen and J. P. Sethna, Phys. Rev.
B\underbar{53} 14872 (1996). 

\item[{[2]}] K. P. O. Brien and M. B. Weissman, Phys. Rev. 
E\underbar{50} 3446 (1994); and references cited therein. 

\item[{[3]}] O. Perkovi\'c, K. A. Dahmen and J. P. Sethna, Cond-mat.
9609072. 

\item[{[4]}] P. Shukla, Physica A \underbar{233}, 235 (1996); Physica A
\underbar{233}, 242 (1996). 

\item [{[5]}] The RFIM on the Bethe lattice with a bivariate quenched
random field and no external uniform field has been studied earlier.  See,
e.g. R. Bruinsma, Phys. Rev. B \underbar{30}, 287 (1984), and references
cited therein. 

\item[{[6]}] D. J. Thouless, Phys.  Rev. Lett. \underbar{56}, 1082 (1986);
ibid. \underbar{57}, 273 (1986). 

\item[{[7]}] K. M. Y. Wong, D. Sherrington, P. Mottishaw, R. Dewar and C.
de Dominicus, J. Phys. A\underbar{21} L99 (1988); E. Akagi, M. Seino and
S. Katsura, Physica A\underbar{178}, 406 (1991); and references cited
therein. 

\item[{[8]}] J. T. Chayes, L. Chayes, J. P. Sethna, and D.  J. Thouless,
Commun.  Math. Phys. 106, 41 (1986); J. M. Carlson, J. T. Chayes, L.
Chayes, J. P. Sethna, and D. J. Thouless, Europhys. Lett. 5, 355 (1988),
J. Stat. Phys. 61, 987 (1990), J. Stat. Phys. 61, 1069 (1990). 

\item[{[9]}] See, for example, K. Kawasaki in {\it Phase Transitions and
Critical Phenomena}, Eds. C. Domb and M.S. Green, Vol. II (Academic,
London, 1972). 

\item[{[10]}] A. A. Middleton, Phys. Rev. Lett. \underbar{68}, 670 (1992). 

\item[{[11]}] R. J. Baxter in {\it Exactly Solved Models in Statistical
Mechanics} (Academic, New York, 1982), pp 47-59. 

\item[{[12]}] L. K. Runnels. J. Math. Phys., \underbar{8}, 2081 (1967). 

\item[{[13]}] P. M. Bleher, J. Ruiz, and V. A. Zagrebnov, J. Stat.  Phys.,
\underbar{84}, 1077 (1996). 

\item[{[14]}] C. Baillee, D. A. Johnston, and J. P. Kownacki, Nucl.  
Phys. 
B\underbar{432}, 557 (1994); C. Baillee, D. A. Johnston, E.  Marinari and
C. Naitza, J. Phys. A\underbar{21}, 6683 (1996). 

\item[{[15]}] B. Bollobas in {\it Random graphs}, (Academic, London,
1985), p53. 

\item[{[16]}] This problem has been studied earlier by S. Maslov and Z. 
Olami, (1993, unpublished).

\item[{[17]}] This value of the coercive field turns out to be dependent
on boundary conditions, and is different if there are some boundary spins
with fewer number of neighbors. 

\end{enumerate}

\newpage

\begin{center} {\large\underbar{\bf Captions to figures}} \end{center}
\bigskip\bigskip

\begin{enumerate} \item[{Fig. 1:}] A Cayley tree of coordination number 3
and height 3. 

\item[{Fig. 2:}] An example of a random graph with coordination number 3. 
Dotted lines indicate the random pairs. 

\item[{Fig. 3:}] Hysteresis loop on the Bethe lattice of coordination
number 3.  Case shown is for standard deviation of quenched random field
$\sigma= J$.  The result of simulation for $N = 10^6$ spins (points) is in
good agreement with our theoretical result (continuous curve). 

\item[{Fig. 4:}] Hysteresis loop on the Bethe lattice for $z=3$ and
$\sigma= 3J$.  The result of simulation for $N = 10^6$ spins (points) is
in good agreement with our theoretical result (continuous curve). 

\item[{Fig. 5:}] Magnetization curves for the Bethe lattice of
coordination number 4 in increasing field. 

\end{enumerate}

\begin{figure}
\centerline{
\epsfxsize=20.0cm
\epsfysize=10.0cm
\epsfbox{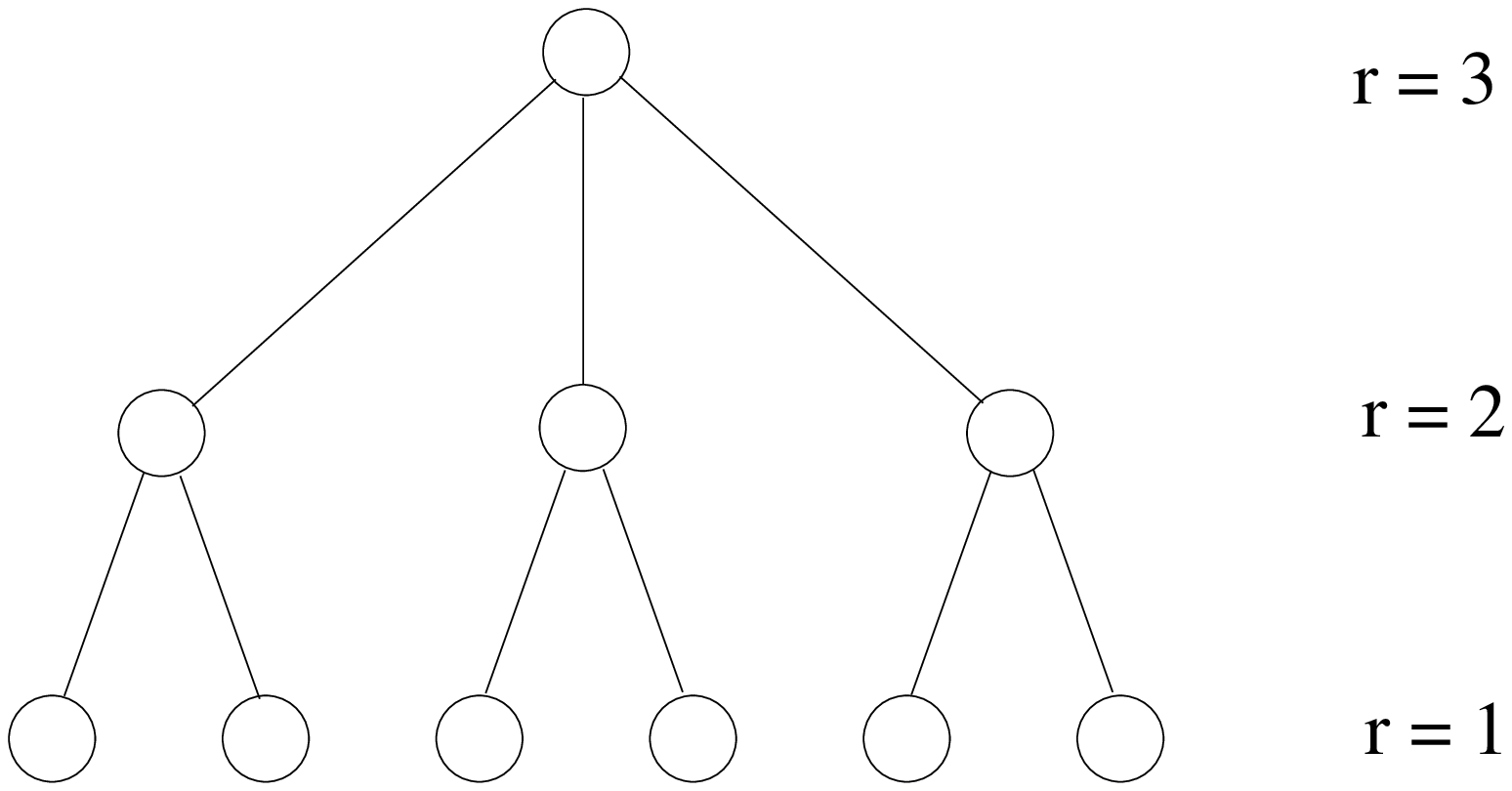}
}
\caption{}
\label{Fig.1}
\end{figure}

\begin{figure}
\centerline{
\epsfxsize=20.0cm
\epsfysize=8.0cm
\epsfbox{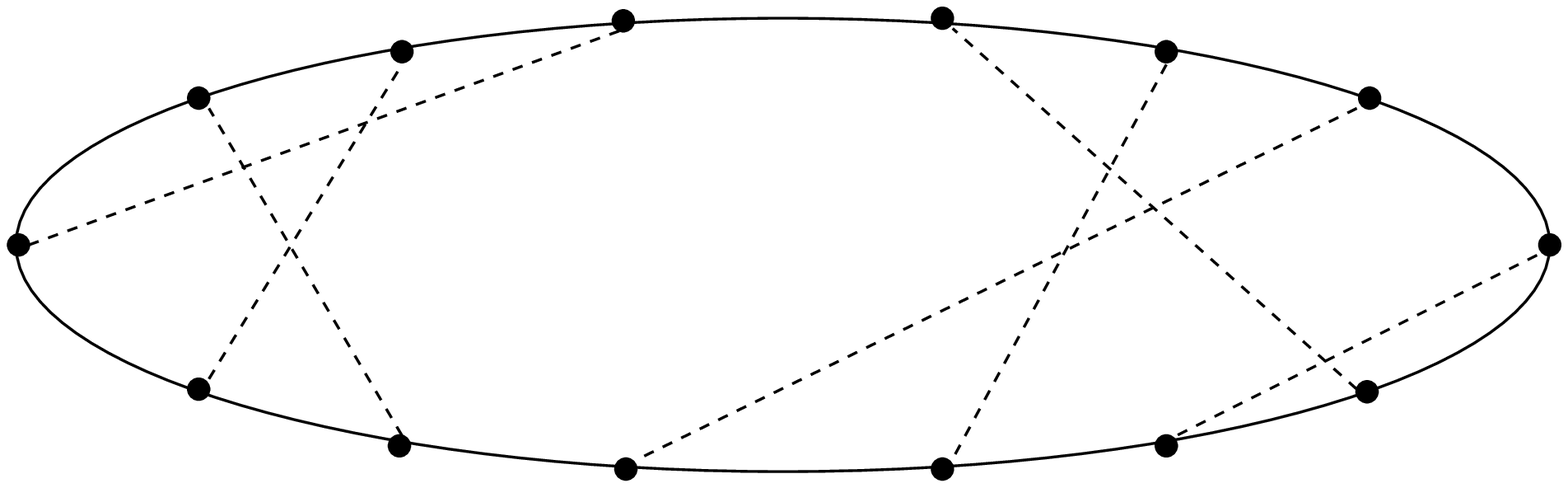}
}
\caption{}
\label{Fig.2}
\end{figure}

\begin{figure}
\centerline{
\epsfxsize=20.0cm
\epsfysize=20.0cm
\epsfbox{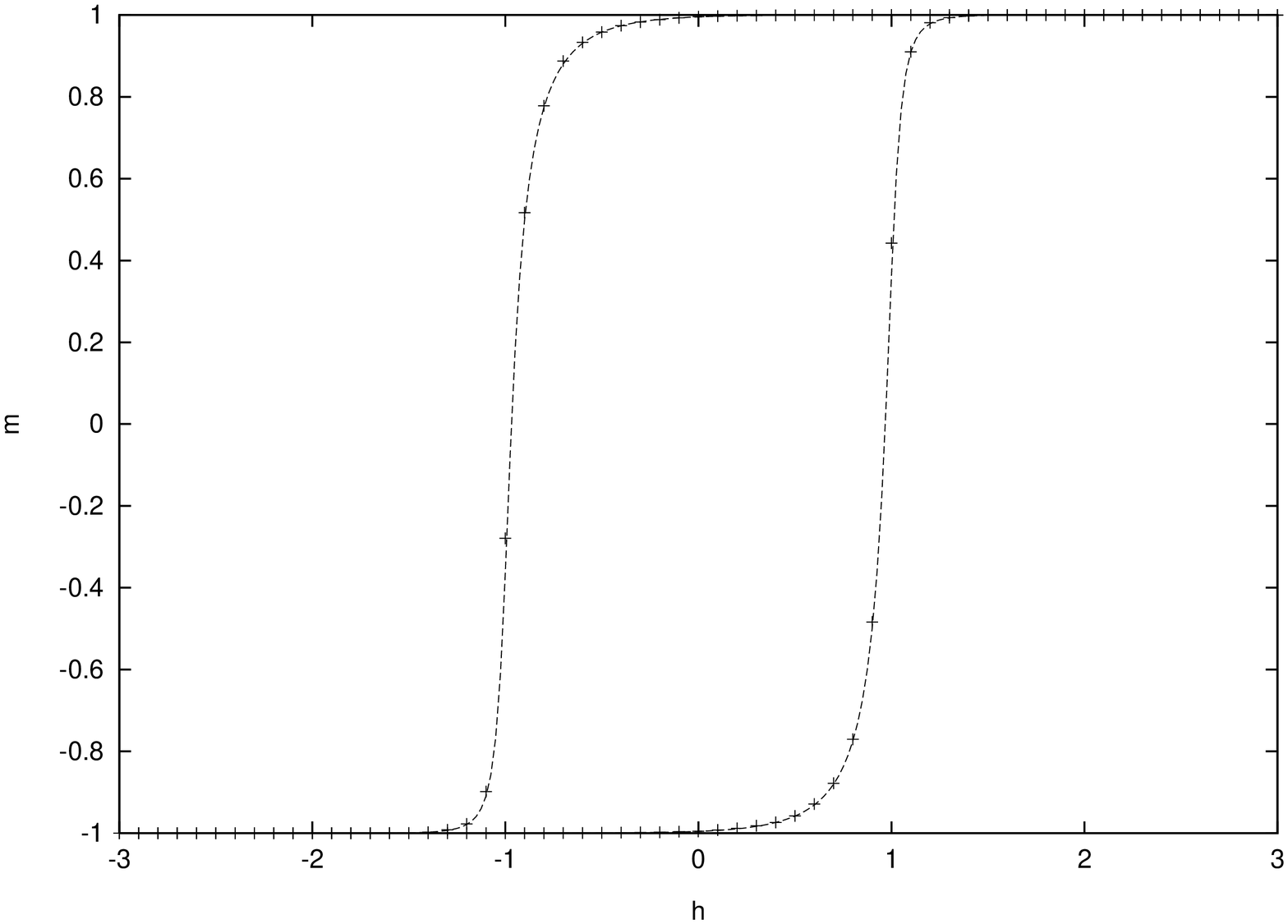}
}
\caption{}
\label{Fig.3}
\end{figure}

\begin{figure}
\centerline{
\epsfxsize=20.0cm
\epsfysize=20.0cm
\epsfbox{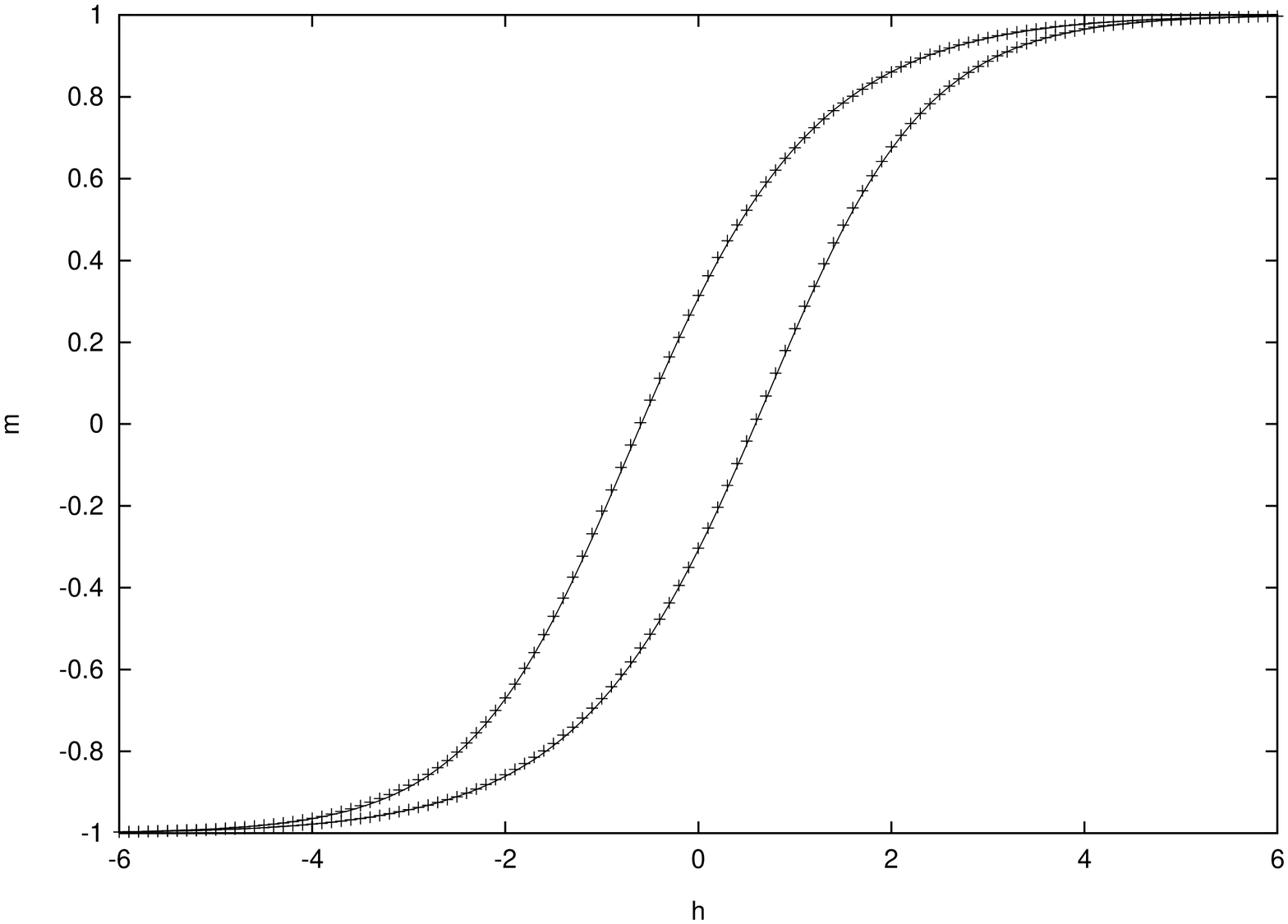}
}
\caption{}
\label{Fig.4}
\end{figure}

\begin{figure}
\centerline{
\epsfxsize=20.0cm
\epsfysize=20.0cm
\epsfbox{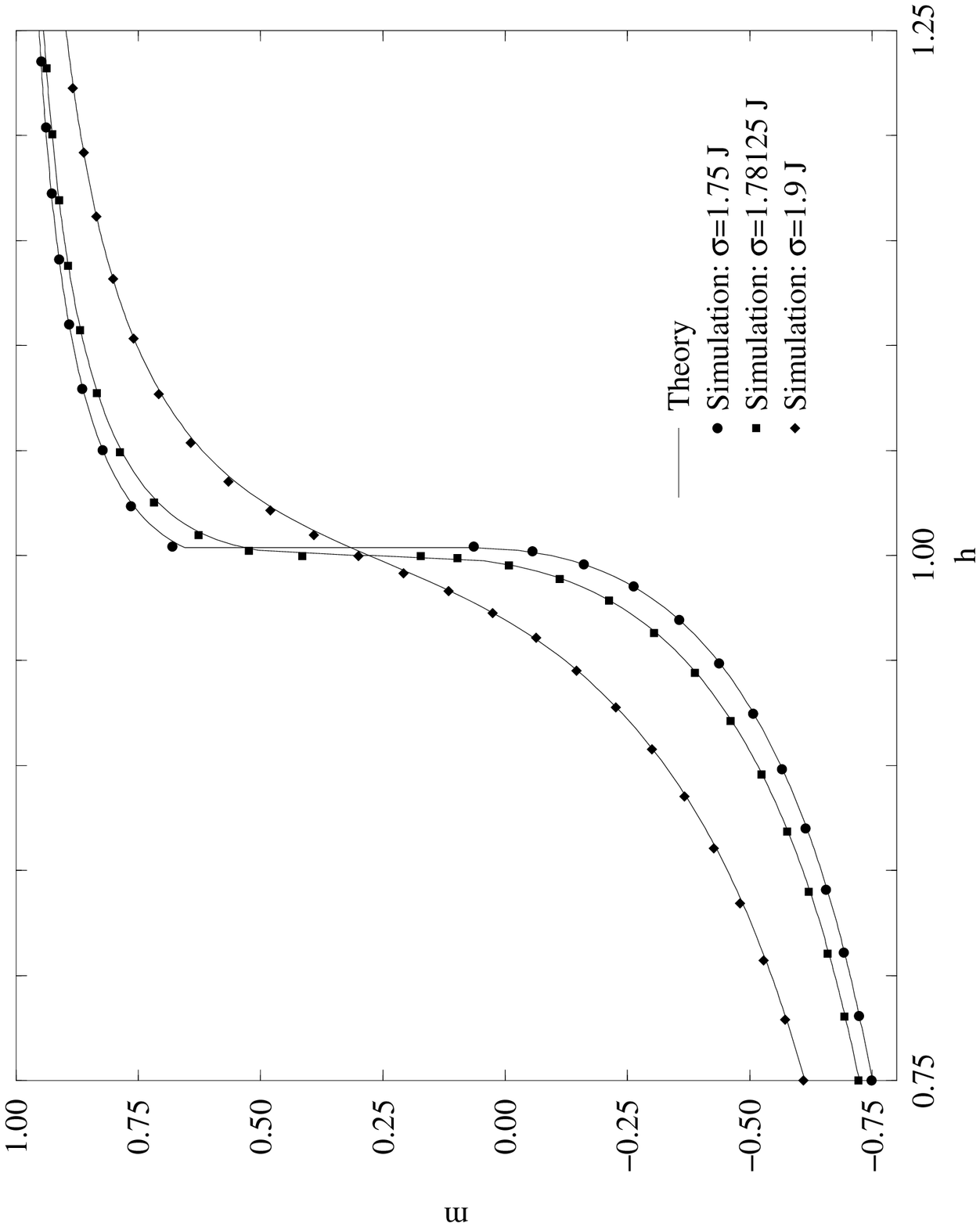}
}
\caption{}
\label{Fig.5}
\end{figure}

\end{document}